\pdfoutput=1
\documentclass[preprint,groupedaddress,nofootinbib]{revtex4-1}

\usepackage{amsfonts}
\usepackage{amsmath}
\usepackage{bm,graphicx}
\usepackage{hyperref}
\usepackage{epsfig}
\usepackage{verbatim}   
\usepackage[subfigure]{graphfig}
\usepackage{epstopdf}
\usepackage{dcolumn}
\usepackage{color}
\usepackage{tabularx}
\usepackage{braket}
\usepackage{float}
\usepackage{tikz}
\usepackage[percent]{overpic}
\graphicspath{{./figures/}}
\usepackage{subfigure}
\usepackage{mwe}
\usepackage{soul}
\usepackage{ulem}

\usepackage[scale=0.8,bmargin=3cm,footnotesep=2cm]{geometry}
\usepackage[flushmargin]{footmisc}

\hypersetup{colorlinks, linkcolor = [rgb]{0,0.0,0.75}, citecolor = [rgb]{0,0.0,0.75}, urlcolor = [rgb]{0,0.0,0.75}}

\usepackage[utf8]{inputenc}
\usepackage{lineno}
\def\be{\begin{equation}}
\def\ee{\end{equation}}
\def\bea{\begin{eqnarray}}
\def\eea{\end{eqnarray}}

\newcommand{\la}{\label}

\definecolor{green}{rgb}{0,.5,0}

\begin{document}

\preprint{SLAC--PUB--16806}

\title{Analysis of nucleon electromagnetic form factors from light-front holographic QCD : The spacelike region }

\author{Raza Sabbir Sufian$^{1}$, Guy F. de T\'eramond$^{2}$, Stanley J. Brodsky$^{3}$,  \\ Alexandre Deur$^{4}$,  Hans G\"{u}nter Dosch$^{5}$ }
\vspace*{-0.5cm}

\affiliation{
$^{1}$\mbox{Department of Physics and Astronomy, University of Kentucky, Lexington, Kentucky 40506, USA}\\
$^{2}$\mbox{Universidad de Costa Rica, 11501 San Jos\'e, Costa Rica}\\
$^{3}$\mbox{SLAC National Accelerator Laboratory, Stanford University, Stanford, California 94309, USA}\\
$^{4}$\mbox{Thomas Jefferson National Accelerator Facility, Newport News, Virginia 23606, USA}\\
$^{5}$\mbox{Institut f\"{u}r Theoretische Physik der Universit\"{a}t, Philosophenweg 16, D-69120 Heidelberg, Germany}\\
}


\begin{abstract}
We present a comprehensive analysis of the spacelike nucleon electromagnetic form factors and their flavor decomposition within the framework of light-front holographic QCD. We show that the inclusion of the higher Fock components $\ket {qqqq\bar{q}}$ has a significant effect on the spin-flip elastic Pauli form factor and almost zero effect on the spin-conserving Dirac form factor. We present light-front holographic QCD results for the proton and neutron form factors  at any momentum transfer range, including  asymptotic predictions, and show that our results agree with the available experimental data with high accuracy. In order to correctly describe the Pauli form factor we need an admixture of a five quark state of about 30$\%$ in the proton and about 40$\%$ in the neutron. We also extract the nucleon charge and magnetic radii and perform a flavor decomposition of the nucleon electromagnetic form factors. The free parameters needed to describe the experimental nucleon form factors are very few: two parameters for the probabilities of higher Fock states for the spin-flip form factor and a phenomenological parameter $r$, required to account for possible SU(6) spin-flavor symmetry breaking effects in the neutron, whereas the Pauli  form factors are normalized to the experimental values of the anomalous magnetic moments. The covariant spin structure for the Dirac and Pauli nucleon form factors  prescribed by AdS$_5$ semiclassical gravity incorporates the correct twist scaling behavior from hard scattering and also leads to vector dominance at low energy.
 
\end{abstract}

\maketitle

\section{Introduction} \la{intro}

The spacelike electromagnetic form factors of the proton and neutron obtained in electron-nucleon elastic scattering are key measures of the fundamental structure of hadrons. The helicity-conserving and helicity-flip current matrix elements required to compute the Dirac $F_1(Q^2) $ and Pauli $F_2(Q^2)$ form factors, respectively, have an exact representation in terms of the overlap of the nonperturbative hadronic light-front wave functions (LFWFs)~\cite{Brodsky:1980zm}, the eigensolutions of the QCD light-front Hamiltonian -- the Drell-Yan-West formulas~\cite{Drell:1969km, West:1970av}. The squares of the same hadronic LFWFs, summed over all Fock states, underly the structure functions measured in deep inelastic lepton-nucleon scattering. A central goal of hadron physics is to not only successfully predict these dynamical observables but to also accurately account for the spectroscopy of hadrons.

The quest for a detailed quantitative understanding of the nucleon form factors is an active field in hadronic physics. A wide variety of models has been proposed to describe the nucleon form factors. However, in most of these approaches there has been no attempt to understand the observed hadron spectroscopy. Furthermore, a consensus among different phenomenological models and parametrizations which describe the nucleon form factors has not yet been achieved, especially for the neutron Dirac and Pauli electromagnetic form factors, and the nucleon time-like form factors.

Detailed reviews of the experimental results and models can be found in Refs.~\cite{Pacetti:2015iqa, Punjabi:2015bba}. It should be noted that inconsistencies in the extraction of the data appear in the proton electric to magnetic Sachs form factor (FF) ratio $R_p(Q^2) =\mu_p G_E^p(Q^2) /G_M^p(Q^2)$, when one compares double polarization experiments~\cite{Gao:2000ne,Punjabi:2005wq,Gayou:2001qd,Puckett:2011xg}, in which the ratio $R_p$ decreases almost linearly for momentum transfer $Q^2>0.5\, \text{GeV}^2$, with the results obtained from the Rosenbluth separation method~\cite{Hand:1963zz,Janssens:1965kd,Price:1971zk,Litt:1969my,Berger:1971kr,Bartel:1973rf,Borkowski:1974mb,Simon:1980hu,Walker:1993vj,Andivahis:1994rq,Christy:2004rc,Qattan:2004ht} in which $R_p$ remains constant in the spacelike (SL) region. Predictions for different combinations of the neutron FFs are even more puzzling to explain using phenomenological models. A further limitation is that experimental data for the neutron FFs are not available in the large $Q^2=-q^2$ regime. Another challenge is to describe the modulus of the electric to magnetic Sachs FF ratio $\vert G_E^p / G_M^p \vert $ measured by the PS170 experiment at LEAR~\cite{Bardin:1994am} and by the {\it{BABAR}} Collaboration in the timelike (TL) domain~\cite{Lees:2013ebn} above the physical threshold $q^2_{phys}=  4 \, m_p^2$, where $m_p$ is the proton mass, at which proton-antiproton pairs are produced at rest in their center of mass system, and where strong threshold effects are also important.

The recent 12 GeV energy upgrade of Jefferson Lab will bring a wealth of high precision measurements at larger $Q^2$. A measurement of $G_M^p$ in Jefferson Lab's Hall A is currently ongoing in the 7 to 17 GeV$^2$ range, with a precision aimed at less than 2 \%~\cite{a}. Future experiments approved for running in Hall A include measurements of $R_p(Q^2)$ up to 15 GeV$^2$ using recoil polarization~\cite{b}, of $R_n(Q^2) =\mu_n G_E^n(Q^2) /G_M^n(Q^2) $ up to 10.2 GeV$^2$ using a polarized $^3$He target~\cite{c}, and of $G_M^n$ up to $Q^2=18$ GeV$^2$ using a deuteron target~\cite{d}. A similar experiment up to $Q^2=14$ GeV$^2$ will run in Jefferson Lab's Hall B~\cite{e}, and a $G_E^n$ measurement up to $Q^2=7$ GeV$^2$ using a deuteron target and recoil polarization will run in Jefferson Lab's Hall C~\cite{f}. Finally, in order to provide an unambiguous value of the proton electric radius from electron scattering, an experiment was recently completed (April 2016) which measured $G_E^p$ down to $Q^2=10^{-4}$ GeV$^2$, with a statistical precision better than $2 \times 10^{-3}$ and a systematic accuracy of $5 \times 10^{-3}$~\cite{g}.

The spectra of hadrons and their FFs can both be calculated using a novel nonperturbative approach to hadron physics called light-front (LF) holographic QCD (LFHQCD)~\cite{Brodsky:2006uqa,deTeramond:2008ht,deTeramond:2013it, Brodsky:2014yha}, which provides new analytical tools for hadron dynamics within a relativistic frame-independent first approximation to the LF QCD Hamiltonian. This new approach to hadronic physics follows from the precise mapping of the Hamiltonian equations in anti-de Sitter (AdS) space to the relativistic semiclassical light-front bound-state equations in the usual Minkowski space~\cite{deTeramond:2008ht,deTeramond:2013it}, which is the boundary space of AdS$_5$. This connection gives an exact relation between the holographic variable $z$ of AdS space and the invariant impact LF variable $\zeta$ in physical space-time \footnote{The invariant light-front variable of an $N-$quark bound state is $\zeta=\sqrt{\frac{x}{1-x}}\big\vert\sum_{j=1}^{N-1}x_j\bf{b}_\perp\big\vert$, where $x$ is the longitudinal momentum fraction of the active quark, $x_j$ with $j =1, 2, \cdots, N-1$, the momentum fractions associated with the $N-1$ quarks in the cluster, and ${\bf{b}}_{\perp j}$ are the transverse positions of the spectator quarks in the cluster relative to the active one~\cite{Brodsky:2006uqa}. For a two-constituent bound-state $\zeta =\sqrt{x(1-x)}\vert{\bf{b}}_\perp\vert$, which is conjugate to the invariant mass $\frac{{\bf{k}}^2_\perp}{x(1-x)}$. }.  This holographic connection also implies that the light-front effective potential $U \sim \kappa^2 \zeta^2$ in the LF Hamiltonian, corresponds to a modification of the infrared region of AdS space. The specific form of the LF potential is determined by superconformal quantum mechanics~\cite{deAlfaro:1976je, Fubini:1984hf, Brodsky:2013ar, deTeramond:2014asa,Dosch:2015nwa}, which captures the relevant aspects of color confinement based on a universal emerging single mass scale $\kappa = \sqrt{\lambda}$~\cite{Brodsky:2016yod}. The  modification is a quadratic dilaton profile in the bosonic AdS$_5$ action and a Yukawa-like interaction term in the fermionic action~\cite{deTeramond:2014asa,Dosch:2015nwa}.

This new approach to hadron physics predicts universal linear Regge trajectories and slopes in both orbital angular momentum and radial excitation quantum numbers, the appearance of a massless pion in the limit of zero-mass quarks, and gives remarkable connections between the light meson and nucleon spectra~\cite{Dosch:2015nwa,Brodsky:2016yod}. The superconformal approach has thus the advantage that mesons and nucleons are treated on the same footing, and the confinement potential is uniquely determined by the formalism. Remarkably, the meson spectrum and baryon spectrum are related by a simple shift of the orbital angular momentum $L_M=L_B+1$. The QCD running coupling is also consistently described at both small and large $Q^2$~\cite{Brodsky:2010ur,Deur:2016cxb,Deur:2016tte}.

In this paper we calculate the spacelike nucleon electromagnetic (EM) form factors within the framework of LFHQCD~\cite{Brodsky:2014yha}. In the modified AdS$_5$ space, which can be considered as a gravity theory in five dimensions, FFs are computed from the overlap integral of normalizable modes, which represent the incoming and outgoing hadrons, convoluted with a non-normalizable mode which represents an  EM current~\cite{Polchinski:2002jw}. The EM current propagates into the infrared modified AdS space and generates an infinite number of poles. Thus, the FF in the gravity theory has the advantage that it generates the nonperturbative pole structure in the timelike region of the FF~\cite{Brodsky:2014yha}. Furthermore, for nucleons, the specific form of the interaction Lagrangian terms in the higher-dimensional gravity theory dictates precise scaling relations for the Dirac and Pauli FFs, which lead, when mapped to physical Minkowski space, to unambiguous scaling predictions for different ratios of nucleon  FFs.

When mapping the ``dressed'' EM current propagating in a modified AdS space to the LF QCD Drell-Yan-West expression for the  FF, the resulting LFWF incorporates nonvalence higher Fock states generated by the confined current~\cite{Brodsky:2011xx}. The gauge/gravity duality also incorporates the connection between the twist-scaling dimension $\tau$ of the QCD boundary interpolating operators with the falloff of the normalizable modes in AdS near its conformal boundary~\cite{Polchinski:2001tt}, consistent with leading-twist scaling; {\it i.e.}, in agreement with the power-law falloff of the counting rules for hard scattering dynamics at large $Q^2$~\cite{Brodsky:1973kr,Matveev:1973ra}. The twist of a particle $\tau$ is defined here as the power behavior of its light-front wave function near $\zeta =0$: $\Phi  \sim \zeta^\tau$. For ground-state hadrons the leading twist is the number of constituents.

When computing nucleon FFs one has to constrain the asymptotic boundary conditions of the leading falloff of the  FFs to match the twist of the nucleon's interpolating operator, {\it i.e.} $\tau = 3$, to represent the fact that at high virtualities the nucleon is essentially a system of three weakly interacting partons. For a multiquark bound state, the LF invariant impact variable $\zeta$ applies to a system composed of an active quark plus a spectator ``cluster". For example, for a three-quark nucleon state, the three-body problem is reduced to an effective two-body problem where two of the constituents form a diquark cluster~\cite{Brodsky:2014yha}. This follows from the holographic approach, where one has only one variable to describe the internal structure of the nucleon. This means, for example, that for a proton the bound state behaves like a quark-diquark system, {\it i.e.}, like a twist-2 system. However, at large momentum transfer, or at small distances, where the cluster is resolved into its individual constituents, the nucleon is governed by twist 3, in contrast to the nonperturbative region where it is approximated by twist 2. A similar feature appears in the study of sequential decay chains in baryons~\cite{TheCBELSA/TAPS:2015ula}, which are sensitive to the short distance behavior of the wave function. At very short distances, the bound state is  twist 3 since the two constituents particles in the diquark are resolved. This different scaling behavior of the structure functions at low and high virtualities can be properly addressed from the LF cluster decomposition for bound states~\cite{Brodsky:1983vf, Brodsky:1985gs, deTeramond:2016pov} and is discussed below.

In contrast to the prototypical example of the gauge/gravity duality, the AdS/CFT correspondence~\cite{Maldacena:1997re}, where the baryon is identified as an $SU(N_C)$ singlet bound state of $N_C$ quarks in the large-$N_C$ limit, in the LFHQCD formalism, baryons are computed for $N_C=3,$ not $N_C\to \infty$. In particular, the correct physical twist assignment is critical when computing hadron FFs since the leading twist corresponds to the number of constituents $N$, {\it i.~e.}, $\tau = 3$ for a nucleon. In fact, the nucleon AdS solutions have both  $L=0$ and $L=1$ components with equal weight. Therefore we use both the leading twist $\tau =3$ and $\tau = 3 +L = 4$ to compute the valence contribution to the nucleon FFs. The spacelike Pauli FF of the nucleons arises from the overlap of $L=0$ and $L=1$ AdS wave functions~\cite{Brodsky:1980zm}. It is important to recall that the spin-flavor symmetry is not contained in the holographic principle, which essentially describes the $Q^2$ scale dependence for a given twist, and has to be imposed from the symmetries of the quark model under consideration. In the present work we use the SU(6) spin-flavor symmetry and examine possible breaking effects of this symmetry.

In holographic QCD gluonic degrees of freedom only arise at high virtuality, whereas gluons with small virtuality are sublimated in the effective confining potential~\cite{Brodsky:2011pw}. Thus, Fock states of hadrons can have any number of extra $q \bar q$ pairs created by the confining potential. One can extend the formalism in order to examine the contribution of higher-Fock states using the holographic framework described here. Indeed, it was shown in Refs.~\cite{Brodsky:2014yha,deTeramond:2010ez} that higher Fock components are essential to describe the rather complex timelike structure of the pion FF. Contribution from the higher-twist components ($q\bar{q}$ and $q\bar{q}q\bar{q}$) has also been considered to describe the pion transition FF in $\gamma\gamma*\to\pi^0$~\cite{Brodsky:2011xx}. Contributions from three, four, and five parton components in the nucleon Fock expansion have been considered in the holographic QCD framework in Ref.~\cite{Gutsche:2012bp}, but the experimental data of a different combination of Sachs FFs, such as $\mu_pG^p_E/G^p_M$, could not be successfully described. More recent works~\cite{Gutsche:2013zia,Gutsche:2014yea} by the same group can describe the experimental data of nucleon FFs well, but the number of parameters required is large, typically about $8 -12$ free parameters. Other attempts to describe the flavor nucleon FFs in AdS/QCD also require a large number of parameters~\cite{Maji:2016yqo}. On the other hand, simple holographic models, which essentially include only the valence contribution, fail to systematically account for all the properties of the nucleon FFs and their flavor decomposition~\cite{Brodsky:2014yha, Chakrabarti:2013dda, Mondal:2016xpk}. As we show below, higher-twist components in the Fock expansion are in general needed for an accurate description of the nucleon FFs, and, in fact, this can be achieved with a minimal number of parameters in the LF holographic framework.

The contents of this article are as follows: After briefly reviewing in Sec. \ref{HFF} how nonperturbative analytical expressions for FFs in physical four-dimensional space follow from semiclassical gravity in AdS$_5$ space, and their light-front holographic cluster decomposition, we show in Sec. \ref{NHFF} how the Dirac and Pauli nucleon FFs in physical space-time follow from the covariant spin structure of FFs in AdS$_5$. In Sec. \ref{NHFFmodel} we study the effect of higher Fock states and build a simple light-front holographic model for the nucleon FFs. We compare our predictions with available data and compute asymptotic predictions for the nucleon FFs and their ratios. We compare our results for the nucleon radii and perform a flavor decomposition of the nucleon FFs. Predictions are made for comparison with upcoming JLab experiments. Our concluding remarks are given in Sec. \ref{conclusions}.

\section{Hadron form factors in holographic QCD} \la{HFF}

For simplicity let us consider first the FF of a spinless hadron. In the higher-dimensional gravity theory an electromagnetic FF corresponds to the coupling of an external  EM field $A^M(x,z)$ propagating in AdS space with a hadron mode $\Phi_P(x,z)$, given by the left-hand side of the equation 
 \begin{multline} \label{FF}
 \int d^4x \, dz \,  \sqrt{g}  \, \Phi_{P'}^*(x,z)  \overleftrightarrow \partial_M \Phi_{P}(x,z)
 \, A^M(x,z)   \sim
 (2 \pi)^4 \delta^4 \left( P'  \! - P - q\right) \epsilon_\mu (P + P')^\mu  F(q^2),
 \end{multline} 
 defined up to a constant term. In (\ref{FF}) the coordinates are $x^M =\left(x^\mu, z\right)$, with $z$ being the holographic variable and $x^\mu$ Minkowski flat space-time coordinates. The metric determinant is $\sqrt{g} = (R/z)^{5}$. To simplify, we set the AdS radius $R=1$ since it does not appear in physical quantities. In the above expression the hadron has initial and final four-momenta $P$ and $P'$ and $q$ is the four-momentum transferred to the hadron by the photon with polarization $\epsilon^\mu$. For convenience we have redefined the wave function $\Phi(x,z)$ to absorb any dependence in Eq. (\ref{FF}) on a dilaton profile. The expression on the right-hand side represents the EM hadron FF in physical space-time. It corresponds to the local coupling of the quark current $J^\mu =  \sum_q e_q \bar q \gamma^\mu q$ to the constituents~\cite{Brodsky:2014yha}.

In holographic QCD a hadron is described by a $z$-dependent wave function which includes the scale dependence and a plane wave in physical space representing a free hadron: $\Phi_P(x,z) = e^{i P \cdot x} \Phi(z)$. The physical incoming electromagnetic probe propagates in AdS according to
\bea
A_\mu(x,z) = e^{i \, q \cdot x} V(q^2,z) \, \epsilon_\mu(q), ~~~~ A_z = 0,
\eea
where the bulk-to-boundary propagator $V(q^2,z)$ has the boundary conditions $V(q^2 = 0, z) = V(q^2, z = 0) = 1$. Extracting the factor $(2 \pi)^4 \delta^4 \left( P'  \! - P - q\right)$ from momentum conservation in Eq. (\ref{FF}) we find~\cite{Polchinski:2002jw} ($Q^2 = - q^2$)
\bea  \label{intFF}
F(Q^2) =  \int  \frac{dz}{z^3} \, V(Q^2, z)  \, \Phi_\tau^2(z),
\eea
where $F(Q^2 = 0)= 0$. At small values of $z \sim 1/Q$, where the EM current $V(Q^2,z)$ has its important support, the hadron modes scale as
$\Phi_\tau \sim z^\tau$, and the hard-scattering power-scaling behavior~\cite{Brodsky:1973kr,Matveev:1973ra} is recovered~\cite{Polchinski:2001tt}
\bea
F(Q^2) \to  \left[\frac{1}{Q^2}\right]^{\tau-1}.
\eea
In our approach the twist-$\tau$ hadronic wave functions are 
\bea \label{Phitau}
\Phi_\tau(z) =   \sqrt{\frac{2 }{\Gamma(\tau \! - \! 1)} } \, \kappa^{\tau -1} z ^{\tau} e^{- \kappa^2 z^2/2},
\eea
and the EM current $V(Q^2,z)$ is the solution of the wave equation of a vector current in AdS$_5$, with modifications determined by the superconformal algebra, which are the same as used in spectroscopy. It has the integral representation~\cite{Grigoryan:2007my} 
\bea \label{Vx}
V(Q^2,z) = \kappa^2  z^2 \int_0^1 \! \frac{dx}{(1-x)^2} \, x^{Q^2 / 4  \kappa^2 }  e^{- \kappa^2  z^2 x/(1-x)}.
\eea
Since the integrand in  Eq. (\ref{Vx}) contains the generating function of the associated Laguerre polynomials $L_n^k$, it can also be expressed as a sum of poles~\cite{Grigoryan:2007my}
\bea \label{Vsum}
V(Q^2, z) = 4 \kappa^4 z^2 \sum_{n=0}^{\infty}  \frac{L_n^1(\kappa^2 z^2)}{ M_n^2 + Q^2},
\eea
with the poles located at $-Q^2 =  M^2_n = 4   \kappa^2  (n + 1)$. To compare with the data, one has, however, to shift the poles in Eq. (\ref{Vsum}) to their physical location at the vector meson masses~\cite{Brodsky:2014yha}
\bea \la{eq6}
-Q^2= M^2_{\rho_n} = 4 \kappa^2 \left(n+ \frac{1}{2} \right), \,\, n=0,1,2, ... \quad .
\eea
The ground-state mass of the $\rho$ meson, $M_{\rho_{n=0}}\equiv M_\rho = 0.775 \,\text{GeV}$ gives the value of $\kappa =   M_\rho / \sqrt 2 = 0.548\, \text{GeV}$, where $\kappa = \sqrt{\lambda}$ is the emerging confinement scale~\cite{Brodsky:2013ar}.

Substituting (\ref{Phitau}) and (\ref{Vx}) in Eq.~(\ref{intFF}), and shifting the vector meson poles to their physical locations using (\ref{eq6}), we find for integer twist  $\tau = N$ the result~\cite{Brodsky:2014yha,deTeramond:2012rt,Brodsky:2007hb}
\bea \label{FFtau} 
  F_{\tau}(Q^2) =  \frac{1}{{\left(1 + \frac{Q^2}{M^2_{\rho_{n=0}}} \right)}
 \left(1 + \frac{Q^2}{M^2_{\rho_{n=1}} } \right) \cdots  \left(1 + \frac{Q^2}{M^2_{\rho_{n = \tau -2}} } \right)},
 \eea
expressed as a product of $\tau -1$ poles along the vector meson Regge radial trajectory in terms of the $\rho$ vector meson mass $M_\rho$ and its radial excitations. For a pion, for example, the leading twist is 2, and thus the corresponding FF has a monopole form~\cite{Brodsky:2007hb}. It is interesting to notice that even if an infinite number of poles appears in the ``dressed'' EM current (\ref{Vsum}), for a twist $\tau$-bound state the corresponding  FF is given by a product of $\tau -1$ poles, thus establishing a precise relation between the twist of each Fock state in a hadron and the number of poles in the hadron FF. As expected from this construction, the analytical form (\ref{FFtau}) incorporates the correct hard-scattering twist-scaling behavior at high virtuality and also vector meson dominance (VMD) at low energy~\cite{Sakurai:1960ju}.

In LF quantization~\cite{Brodsky:1997de}, a hadron state $\vert H \rangle$ is a superposition of an infinite number of Fock components $\vert N \rangle$, $\vert H \rangle = \sum_N \psi_{N/H} \vert N \rangle$, where $\psi_{N/H}$ represents the $N$-component LFWF with normalization $\sum_N \vert \psi_{N/H} \vert^2 = 1$. Thus the FF is given by the sum over an infinite number of terms
\begin{equation} \label{FFtausum}
F_H(Q^2) = \sum_\tau P_\tau F_\tau(Q^2),
\end{equation}
where $F_\tau$ is given by Eq.~(\ref{FFtau}). Since the charge is a diagonal operator, only amplitudes with an identical number of components in the initial and final states contribute to the sum in Eq. (\ref{FFtausum}). Normalization at $Q^2 =0$, $F_H(0) = 1$, $F_\tau(0)=1$~[Eq. (\ref{FFtau})] implies that $\sum_\tau P_\tau = 1$ if all possible states are included.

Conventionally, the analysis of FFs is based on the generalized vector meson dominance model where the EM form factor is written as a single-pole expansion 
 \begin{equation} \label{FFlambda}
F_H(Q^2) = \sum_\lambda C_\lambda \frac{M^2_\lambda}{M^2_\lambda - Q^2},
\end{equation}
with a dominant contribution from the $\rho$ vector meson plus contributions from the higher resonances $\rho'$, $\rho''$, $\rho'''$, \dots, etc.~\cite{Kuhn:1990ad}. The comparison of Eqs.~(\ref{FFtausum}) and~(\ref{FFlambda}) allows us to determine the coefficients $C_\lambda$ in terms of the probabilities $P_\tau$ for each Fock state and the vector meson masses $M_{\rho_n}^2$. The advantage, however, of the holographic approach is that no fine-tuning of the coefficients $C_\lambda$ is necessary since the correct scaling is incorporated from the onset; the expansion coefficients $P_\tau$ then have a clear physical meaning in terms of the probability of each Fock component.

The expression for the FF (\ref{FFtau}) contains a cluster decomposition: the hadronic FF factorizes into the $i = N - 1$ product of twist-2 monopole FFs evaluated at different scales~\cite{deTeramond:2016pov} ($N$ is the total number of constituents of a given Fock state)
 \bea
F_{i }(Q^2) = F_{i = 2}\left(Q^2\right)  \,F_{i = 2} \left(\tfrac{1}{3}Q^2\right)\,  \cdots\,    F_{i = 2} \left(\tfrac{1}{ 2 i - 1} Q^2\right).
\eea
In the case of a nucleon, for example, the Dirac FF of the twist-3 valence quark-diquark state $F_1(Q^2)  = F_{i = 2}\left(Q^2\right)\,  F_{i=2}  \left(\tfrac{1}{3}Q^2\right)$ corresponds to the factorization of the proton FF as a product of a pointlike quark and a diquark-cluster FF. The identical twist-3 expression from Eq. (\ref{FFtau}) is described by the product of two poles consistent with leading-twist scaling, $Q^4 F_1(Q^2) \sim  \text{const}$, at high momentum transfer.  As we will show below, the Pauli form factor $F_2$ is given instead by the $i = N + 1$ product of dipoles, and thus the leading-twist scaling $Q^6 F_2(Q^2) \sim  \text{const}$.

\section{Nucleon Form Factors in Holographic QCD} \la{NHFF}

The nucleon spin-non-flip EMFF follows from the expression~\cite{Brodsky:2014yha}
 \begin{multline} \label{FFD}
 \int d^4x \, dz \,  \sqrt{g}   \,  \bar\Psi_{P'}(x,z)
 \,  e^M_A  \, \Gamma^A \, A_M(x,z) \Psi_{P}(x,z) \\  \sim 
 (2 \pi)^4 \delta^4 \left( P'  \! - P - q\right) \epsilon_\mu \bar u(P') \gamma^\mu F_1(q^2) u({P}),
 \end{multline} 
where the curved space indices in AdS$_5$ space are $M, N$, and tangent indices in flat five dimensional space are $A, B$. The $\Gamma_A$ are Dirac gamma matrices which obey the usual anticommutation relation $\{\Gamma^A, \Gamma^B\} = 2 \eta^{AB}$ and are given by $\Gamma^A = (\gamma^\mu, - i \gamma^5)$, and the $e^M_A$ are  the inverse vielbein, $e^M_A = \left(\frac{z}{R}\right) \delta^M_A$. The expression on the right-hand side represents the Dirac EM form factor in physical space-time. It is the EM spin-conserving matrix element of the quark current $J^\mu = \sum_q e_q \bar q \gamma^\mu q$~\cite{Brodsky:2014yha}.

In the higher-dimensional gravity theory nucleons are described by plus and minus wave functions $\Psi_+$ and $\Psi_-$ corresponding to the positive and negative chirality of the nucleon~\cite{deTeramond:2013it,Brodsky:2014yha}
\bea \label{Psipm}
 \Psi_+(z) \sim  z ^{\tau+1/2} e^{- \kappa^2 z^2/2},  ~~~~~~
 \Psi_-(z)  \sim z ^{\tau+3/2} e^{- \kappa^2 z^2/2},
 \eea
which represent, respectively,  a positive chirality component with orbital angular momentum $L=0$ and and a negative chirality component with $L=1$, and have identical normalization. The spin-nonflip nucleon elastic form factor $F_1$ follows from (\ref{FFD}) and is given in terms of $\Psi_+$  and $\Psi_-$~\cite{Brodsky:2014yha},
\bea \label{FFDpm} 
F_1^N(Q^2) = \sum_\pm g^N_{\pm}  \int  \frac{dz}{z^4} \, V(Q^2, z)  \, \Psi_\pm^2(z).
\eea
 The effective charges $g_\pm$ have to be determined by the specific spin-flavor structure which is not contained in the holographic  principle. For example, in the SU(6) symmetry approximation the effective charges are computed by the sum of the EM charges of the struck quark convoluted by the corresponding probability for the $L=0$ and $L=1$ components $\Psi_+$ and $\Psi_-$ respectively. The result is~\cite{Brodsky:2014yha}
\bea \label{Neffch}
g_+^p = 1,   ~~~~  g_-^p = 0,    ~~~~ g_+^n = - \frac{1}{3}, ~~~~  g_-^n = \frac{1}{3}.
\eea
Notice that there is an additional scaling power  in (\ref{FFDpm}), as compared with Eq. (\ref{intFF}), but this is compensated by the additional $z^{1/2}$ factor in the twist-$\tau$ nucleon AdS wavefunctions (\ref{Psipm}).

Since the structure of  (\ref{FFD}) can only account for $F_1$, one should therefore include an effective gauge-invariant interaction in the five-dimensional gravity action to describe the spin-flip amplitude~\cite{Abidin:2009hr}. The nucleon spin-flip EMFF follows from the nonminimal term
 \begin{multline} \label{FFP}
 \int d^4x \, dz \,  \sqrt{g}   \,  \bar\Psi_{P'}(x,z)
 \,  e^M_A   \,  e^N_B \, \left[\Gamma^A,  \Gamma^B \right]\, F_{M N}(x,z) \, \Psi_{P}(x,z) \\  \sim 
 (2 \pi)^4 \delta^4 \left( P'  \! - P - q\right) \epsilon_\mu \bar u(P') \, \frac{\sigma^{\mu \nu} q_\nu}{2 M_N} \, F_2(q^2) u({P}),
 \end{multline} 
where the expression on the right-hand side represents the Pauli EM form factor in physical space-time. It corresponds to the EM spin-flip matrix element of the quark current $J^\mu = \sum_q e_q \bar q \gamma^\mu q$~\cite{Brodsky:2014yha}. Since (\ref{FFP}) represents an effective interaction, its overall strength has to be fixed to the static values of the anomalous magnetic moments $\chi_p$ and $\chi_n$~\cite{Brodsky:2014yha, Abidin:2009hr}.

Extracting the factor $(2 \pi)^4 \delta^4 \left( P'  \! - P - q\right)$ from momentum conservation in (\ref{FFP}) we find~\cite{Abidin:2009hr} 
\bea \label{FFPpm} 
 F_2^N(Q^2) = \chi_N  \int  \frac{dz}{z^3} \, \Psi_+(z) \,V(Q^2, z)  \, \Psi_-(z), 
\eea
where $N = p,n$. Comparing the spin-flip result (\ref{FFPpm}) with the with the spin-nonflip FF (\ref{FFDpm}), it becomes clear that there is an important difference between the scaling powers of $z$ in both expressions. This difference arises from two sources: first, the appearance of one vielbein in (\ref{FFD}), but two in Eq. (\ref{FFP}), and secondly, the appearance of an additional power of $z$ in the product of the two wave functions due to the different scaling behavior of $\Psi_+$ and $\Psi_-$, Eq. (\ref{Psipm}), with orbital angular momentum $L=0$ and $L=1$ respectively. As a result, while the leading scaling behavior of the Dirac form factor is $1/Q^4$, the leading scaling behavior of the Pauli form factor is $1/Q^6$ because of the additional $z^2$-factor in (\ref{FFPpm}). Remarkably, the correct large-$Q^2$ power scaling from hard scattering is incorporated in the covariant spin structure of the AdS expressions for the nucleon FFs.

\section{A Simple Light-Front Holographic Model for Nucleon Form Factors} \label{NHFFmodel}

Following Ref.~\cite{deTeramond:2010ez} we will consider a simplified model where we only include the first two components in a Fock expansion of the nucleon LF function with no constituent dynamical gluons~\cite{Brodsky:2011pw}
\bea 
\vert N \rangle_{L=0}  &= & \psi^{L=0}_{q q q / N} \vert q q q   \rangle_{\tau=3} 
+  \psi^{L=0}_{q q q q \bar q /N} \vert q  q  q q \bar q  \rangle_{\tau=5}  +   \cdots,  \\
\vert N \rangle_{L=1}   & = & \psi^{L=1}_{q q q / N} \vert q q q   \rangle_{\tau=4} 
+  \psi^{L=1}_{q q q q \bar q/N} \vert q q q  q \bar q  \rangle_{\tau=6}  +   \cdots,
\eea
with $N = p,n$. The additional $q \bar q$ contribution to the nucleon wave function from higher Fock components is relevant at larger distances and is usually interpreted as a pion cloud.

We have performed a systematic evaluation of the relevance of higher Fock components in the nucleon FFs by extending the previous results in Ref.~\cite{Brodsky:2014yha} for the Dirac and Pauli FFs. For example, for the proton Dirac FF we have determined the relevance of higher Fock components by writing $F_1^p(Q^2) = (1-\alpha_p)\,F_{i=3}(Q^2) + \alpha_p \, F_{i=5}(Q^2)$, where $i - 1$ is the number of poles in the expansion (\ref{FFtau}) and $\alpha_p$ is the twist-5 probability $\alpha_p = P^\alpha_{q  q  q q \bar q/p}$. Therefore, $1 - \alpha_p = P^\alpha_{q  q  q/p}$ is the valence twist-3 probability for the spin non-flip EM transition amplitude. It is found that $ P_{q  q  q q \bar q/p}$ is very small, of the order of {\mbox {1 \%}}. Likewise, the contribution of higher Fock components to the Dirac neutron FF is of the order of 2 \% and does not change significantly our previous results~\cite{Brodsky:2014yha}. We thus drop the contribution of the higher Fock components to the spin-nonflip nucleon FFs in the rest of our analysis;  namely, we take $P^\alpha_{q  q  q/p} = P^\alpha_{q  q  q/n } = 1$, which gives us a considerable simplification. Within this approximation, thus considering only the effect of higher $q \bar q$ Fock components to the spin-flip nucleon FFs, we write 
\bea \label{protonF1} 
F_1^p(Q^2) &=& F_{i=3}(Q^2),  \\  \la{protonF2}
F_2^p(Q^2) &=& \chi_p [(1-\gamma_p)F_{i=4}(Q^2) + \gamma_p F_{i=6}(Q^2)]
\eea
for the proton,  where $\chi_p=\mu_p-1=1.793$ is the proton anomalous moment, and 
\bea \la{neutronF1}
F_1^n(Q^2) &=&  -\frac{1}{3}\left[F_{i=3}(Q^2) - F_{i=4}(Q^2) \right] ,  \\ \la{neutronF2} 
F_2^n(Q^2) &=& \chi_n \left[(1- \gamma_n)F_{i=4}(Q^2) + \gamma_n F_{i=6}(Q^2) \right]
\eea
for the neutron, with $\chi_n=\mu_n=-1.913$, and where $\gamma_{p,n}$ are the higher Fock probabilities for the $L=0 \to L=1$ spin-flip nucleon EM form factors. Equations (\ref{protonF1}) and (\ref{neutronF1}) are the exact SU(6) results for the spin-nonflip nucleon FFs  (\ref{FFD}) in the valence configuration, whereas (\ref{protonF2}) and (\ref{neutronF2}) correspond to the spin-flip nucleon FFs  (\ref{FFP}), incorporating the higher Fock components, properly normalized to the nucleon anomalous magnetic moments.

The inclusion of higher Fock states is not of much help in describing well the available data for the neutron Dirac form factor. Indeed the zero value of the neutron FF at zero momentum transfer comes from the cancellation of two normalizable wave functions, which vanishes at $Q^2=0$. One could thus expect that in contrast to the other three FFs, namely $F_{1}^{p}$, $F_{2}^{p}$ and $F_{2}^{n}$, second order effects are more important. Therefore our results for the neutron FFs are, in principle, less reliable than our predictions for the proton FFs, especially for the low $Q^2$ region which is more sensitive to the leading cancellations. With this possible shortcoming of the model in mind, we are thus led to introduce one additional parameter $r$, which is required phenomenologically. With this free parameter $r$ we modify the neutron effective charges in Eq. (\ref{Neffch}) as
\bea \label{neffch}
g_+^n = - \frac{1}{3} r, ~~~~~~  g_-^n = \frac{1}{3} r,
\eea
and thus the expression for the neutron Dirac FF
\bea \la{F1nmod}
F_1^n(Q^2) =-\frac{1}{3}  r \, [F_{\tau=3}(Q^2) - F_{\tau=4}(Q^2) ].
\eea
The value $r=2.08$ is required to give a proper matching
to the available experimental data as shown in Fig.~\ref{fig:q4f1}. Also,
keeping in mind that the gauge-gravity duality does not determine
the spin-flavor structure of the nucleons, which is conventionally included in the
nucleon wave function using SU(6) spin-flavor symmetry, the departure of this free parameter $r$ from unity may be interpreted as a
SU(6) symmetry-breaking effect in the neutron Dirac FF. Indeed, the breaking of SU(6) flavor-spin symmetry
has also been observed in a meson cloud model where mixed symmetry in the
nucleon wave function was included to reproduce the experimental data~\cite{Pasquini:2007iz}.
The effect of SU(6) symmetry breaking on the neutron FFs was also
investigated within a LF constituent quark model in Ref.~\cite{Cardarelli:2000tk}.

\begin{figure}[htp]
\begin{center}
\includegraphics[height=7.0cm,width=10.5cm]{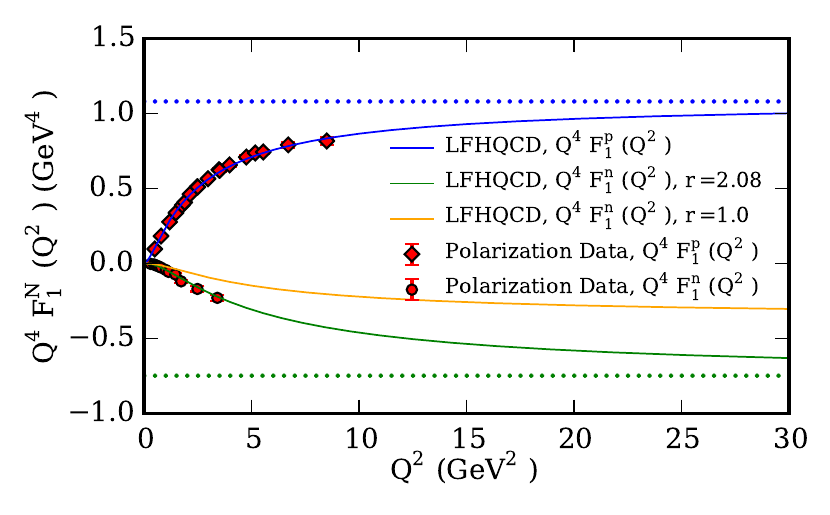}
\end{center}
\setlength\abovecaptionskip{-9pt}
\setlength\belowcaptionskip{-5pt}
\caption{Polarization measurements and predictions for the proton and neutron Dirac form factors~\cite{Riordan:2010id,Qattan:2012zf}. The blue line is the prediction of the proton Dirac FF from LFHQCD, Eq.~(\ref{protonF1}) multiplied by $Q^4$. The orange and the green lines are predictions for the neutron Dirac FF, $Q^4F_1^n(Q^2)$, from Eq.~(\ref{neutronF1}) and from Eq.~(\ref{F1nmod}) with the phenomenological factor $r=2.08$, respectively. The dotted lines are the asymptotic predictions. The asymptotic value of the neutron FF is determined using $r=2.08$.}
\la{fig:q4f1}
\end{figure}

\begin{figure}[htp]
\begin{center}
\includegraphics[height=7.0cm,width=10.5cm]{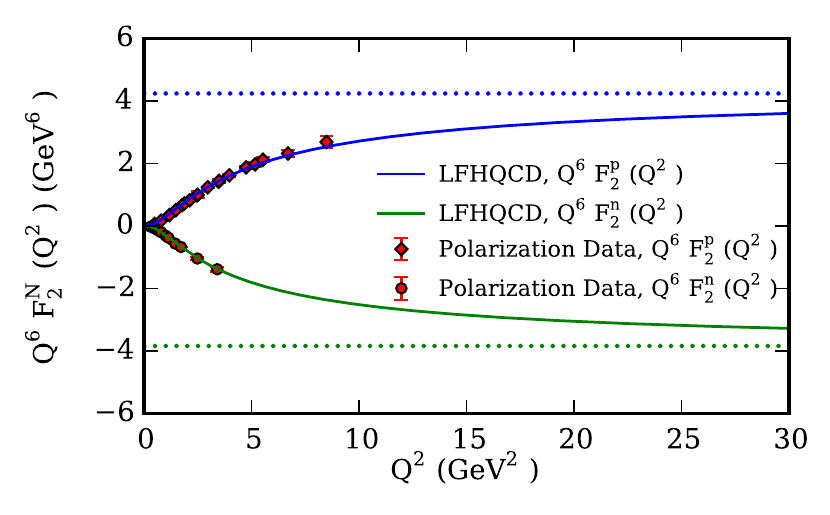}
\end{center}
\setlength\abovecaptionskip{-9pt}
\setlength\belowcaptionskip{-5pt}
\caption{Polarization measurements and predictions for the proton and neutron Pauli form factors~\cite{Riordan:2010id,Qattan:2012zf}. The blue line is the proton Pauli FF, $Q^6F_2^p(Q^2)$ prediction, with $\gamma_p=0.27$ in Eq.~(\ref{protonF2}). The green line is the prediction for the neutron Pauli FF, $Q^6F_2^n(Q^2)$, with $\gamma_n=0.38$ in Eq.~(\ref{neutronF2}) from LFHQCD. The dotted lines are the asymptotic predictions.} 
\la{fig:q6f2}
\end{figure}

All the results presented here correspond to the value of the universal confinement scale determined from the mass of the rho vector meson: $\kappa  =\sqrt \lambda = m_\rho/ \sqrt 2 = 0.548$ GeV. We estimate the uncertainties in our predictions from the uncertainty of the confinement scale $\kappa$. The universality of $\kappa$ is affected -- typically at the $10\%$ level~\cite{Brodsky:2016yod} -- by the inherent approximations of the LFHQCD model~\cite{Brodsky:2014yha}. We discuss the estimate of the model uncertainties in  Appendix \ref{appendix}.

From Figs.~\ref{fig:q4f1} and ~\ref{fig:q6f2}, it is evident that the contribution of an additional $q\bar{q}$ pair, which embodies the pion cloud in the nucleon, only plays an important role in reproducing the experimental data for the spin-flip Pauli FFs. Such an effect of the pion cloud has been addressed in various calculations, for example in Ref.~\cite{Cloet:2012cy}, to show that the same light-front model fails to reproduce the neutron electric Sachs FF $G^n_E$, unless the effect of the pion cloud is included. An estimate reported in Ref.~\cite{Cloet:2014rja} indicates that the pion loop effect results in a $6\%$ and $12\%$ increase in proton charge and magnetic radii, respectively. For the neutron, the effects are a $65\%$ and a $19\%$ increase in charge and magnetic radii, respectively. The values of $\gamma_p$ and $\gamma_n$ show that the effect of the pion cloud on the Pauli FF is larger for the neutron. The dotted lines in Figs.~\ref{fig:q4f1} and ~\ref{fig:q6f2} are the asymptotic results for the nucleon Dirac and Pauli FFs determined by LFHQCD consistent with the QCD power-counting rules. The asymptotic value of the FFs can be easily obtained from (\ref{eq6}) and (\ref{FFtau}). We obtain directly
\begin{equation} \label{asymptotic} 
\lim_{Q^2\to \infty}  \left(Q^2\right)^{\tau -1}
F_\tau(Q^2) = M_{n=0}^2  \cdots  M^2_{n= \tau-2}=\kappa^{2 \tau -2} \prod_{n=0}^{\tau -2} (2 + 4n).
\end{equation}

In the large-$Q^2$ domain the power counting rules are reproduced by the model which also determines its asymptotic normalization. For the spin-nonflip EM nucleon FFs we obtain from (\ref{protonF1}), (\ref{F1nmod}) and (\ref{asymptotic}) the asymptotic results 
\begin{equation} 
\lim_{Q^2\to \infty}  Q^4
F^p_1(Q^2) = M_{n=0}^2 \,  M_{n=1}^2 = 12\,\kappa^4,
\end{equation}
and
\begin{equation} 
\lim_{Q^2\to \infty}  Q^4
F^n_1(Q^2) = - \frac{1}{3} \, r \, M_{n=0}^2 \,  M_{n=1}^2 = - 4 r\,\kappa^4,
\end{equation}
since the valence probability $P^\alpha_{qqq / p} \simeq P^\alpha_{qqq / n}  \simeq1$. On the other hand, for the spin-flip EM nucleon FFs we obtain
\begin{equation} 
\lim_{Q^2\to \infty}  Q^6
F^N_2(Q^2) = \chi_N P^\gamma_{qqq / N} M_{n=0}^2 \,  M_{n=1}^2 \,  M_{n=2}^2= 120\,\chi_N P^\gamma_{qqq / N} \, \kappa^6,
\end{equation}
where $P^\gamma_{qqq / N} = (1 - \gamma_N)$, $N = p, n$, is the valence probability for the spin-flip EM transition amplitude. Possible logarithmic corrections are, of course, not predicted in this semiclassical model.

Another pair of FFs, called the electric and the magnetic Sachs FFs are defined by a combination of Dirac and Pauli FFs as follows:
\bea 
G^{N}_E(Q^2) &=& F_1^{N}(Q^2)-\frac{Q^2}{4m_N^2}F_2^{N}(Q^2),\\ 
G^{N}_M(Q^2) &=&  F_1^{N}(Q^2) + F_2^{N}(Q^2) .
\eea

The results of the ratio $R_p=\mu_p G_E^p/G_M^p$ from the polarization experiments have triggered a revision of various nucleon models, and for $Q^2>10\, \text{GeV}^2$ the ratio $R_p$ may vanish or become negative. We present in Fig.~\ref{fig:Rp} the LFHQCD prediction of $R_p$ up to $Q^2=30\, \text{GeV}^2$, and compare our result with selected world data of unpolarized cross section and polarization measurement experiments. It is clearly seen from Fig.~\ref{fig:Rp} that LFHQCD predicts that $G_E^p$ will decrease more rapidly than $G_M^p$ for $Q^2> 1\, \text{GeV}^2$, in agreement with the polarization measurements of $R_p$. The asymptotic result for $R_p$ follows from
\begin{equation}
\lim_{Q^2\to \infty} 
R_p(Q^2) =  \mu_p \left(1- \frac{5}{2}(\mu_p-1) P^\gamma_{qqq/p} \, \frac{\kappa^2}{m_p^2}\right),
\end{equation}
and has the value $R_p(\infty) = - 0.309$ as indicated in Fig.~(\ref{fig:Rp}). The monotonic decrease of $R_p$ with $Q^2$ demonstrates that the FFs are not simply the sum of dipolelike contributions from the up and down quarks. Following the discussion presented in Appendix \ref{appendix}, we have included in Fig.~\ref{fig:Rp} an estimate of uncertainties in the LFHQCD model. The uncertainty band has been presented with a smooth transition between nonperturbative and perturbative estimates near the transition point $Q_0^2 \simeq 1.5\,\text{GeV}^2$.

\begin{figure}[htp]
\begin{center}
\includegraphics[height=7.0cm,width=10.5cm]{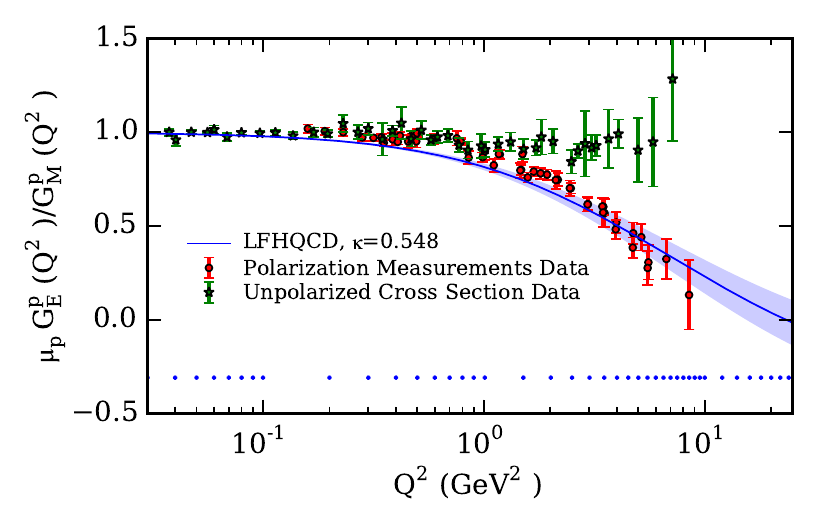}
\end{center}
\setlength\abovecaptionskip{-9pt}
\setlength\belowcaptionskip{-5pt}
\caption{LFHQCD prediction and comparison with selected world data of the ratio $R_p=\mu_p G_E^p/G_M^p$ from unpolarized cross section measurements from~\cite{Bartel:1973rf,Price:1971zk,Hanson:1973vf,Borkowski:1974mb} and polarization measurements from~\cite{Jones:1999rz,Punjabi:2005wq,Gayou:2001qd,Puckett:2010ac,Jones:2006kf,Ron:2011rd,Crawford:2006rz}. The LFHQCD prediction (blue line) from Eqs. (\ref{protonF1}) and (\ref{protonF2}) corresponds to the range $0\,  \leq Q^2 \leq 30\, \text{GeV}^2$. The band represents an estimated theoretical uncertainty of the model. Our theoretical results agree well with the polarization data and are incompatible with the experimental results obtained from Rosenbluth separation. The dotted line is the asymptotic prediction $R_p(\infty)=-0.309$ with an estimated uncertainty of $\pm 0.12$.}
\la{fig:Rp}
\end{figure}

In contrast to the proton FFs, the neutron FFs are more difficult to measure because there is no free neutron target. Experimental data of neutron FFs are available only up to relatively small values of $Q^2$. Since most nucleon form factor models such as~\cite{Miller:2002ig,Miller:2002qb,Cardarelli:2000tk,Lomon:2002jx} cannot reproduce the experimental data for the ratio $R_n=\mu_nG_E^n /G_M^n$ for $Q^2 \geq 2 \, \text{GeV}^2$, it is desirable that one can parametrize the ratio $R_n$ according to the available experimental data and predict its behavior at large $Q^2$. To this end, we compare in Figs.~\ref{fig:GnE} and \ref{fig:Rn} the Sachs electric FF and the ratio $R_n$, computed in LFHQCD, with selected experimental data. From these results, one can see that LFHQCD can properly reproduce $G^n_E$ and $R_n$ in the whole range of available experimental data. We have also extended our results for the neutron FFs to higher $Q^2$ in order to compare with upcoming JLab experiments~\cite{c,d,e,f}. Here the asymptotic value depends in a nontrivial way on the parameter $r$,
\begin{equation}
\lim_{Q^2\to \infty} 
R_n(Q^2)= \mu_n \left(1+\frac{15 \, \mu_n P^\gamma_{qqq/n}}{2 r} \frac {\kappa^2}{m_n^2} \right),
\end{equation}
and has the value $R_n(\infty) = 0.864$ for $r = 2.08$ as indicated in Fig.~(\ref{fig:Rn}).

\begin{figure}[htp]
\includegraphics[height=7.0cm,width=10.5cm]{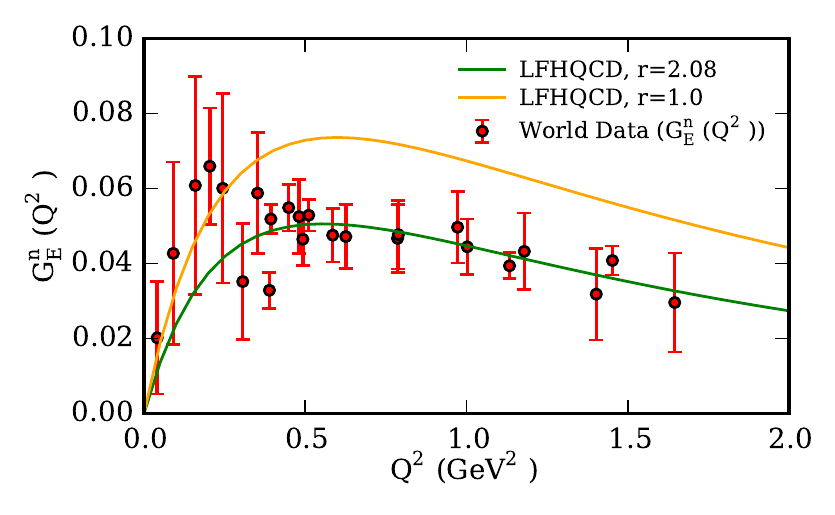}\hfill
\includegraphics[height=7.0cm,width=10.5cm]{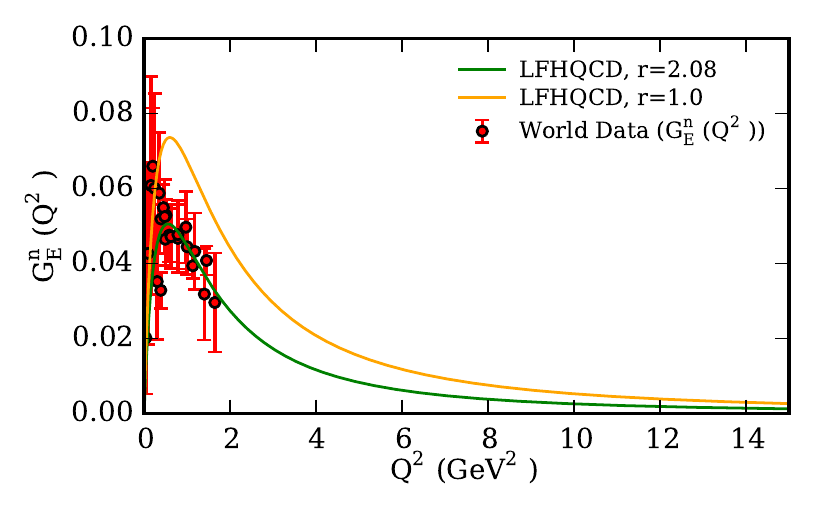}\hfill
\caption{Comparison of the neutron electric FF $G^n_E(Q^2)$ world data~\cite{Glazier:2004ny,Herberg:1999ud,Becker:1999tw,Meyerhoff:1994ev,Golak:2000nt,Passchier:1999cj,Schiavilla:2001qe,Madey:2003av,Zhu:2001md,Warren:2003ma,Plaster:2005cx} with the LFHQCD prediction from Eqs.~(\ref{neutronF1}), (\ref{neutronF2}) and~(\ref{F1nmod}).}
\la{fig:GnE}
\end{figure}

\begin{figure}[htp]
\begin{center}
\includegraphics[height=7.0cm,width=10.5cm]{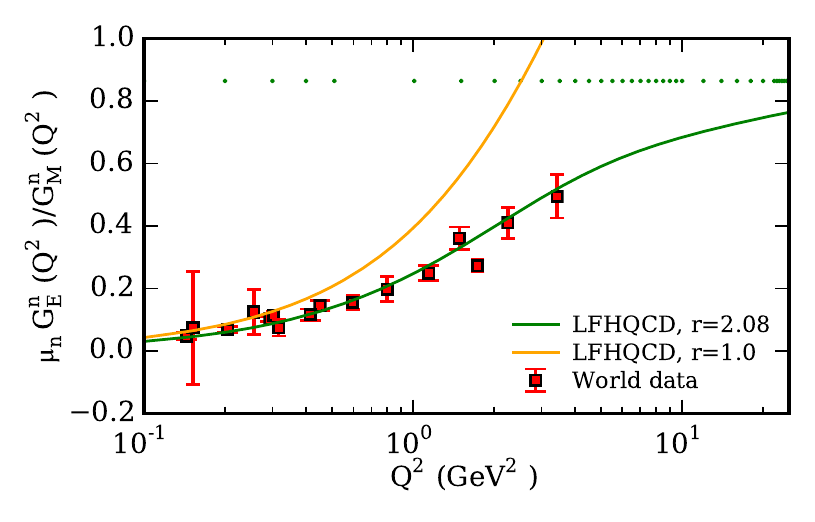}
\end{center}
\setlength\abovecaptionskip{-9pt}
\setlength\belowcaptionskip{-5pt}
\caption{Selected world data of the ratio $R_n = \mu_nG_E^n /G_M^n$ from double polarization experiments; recoil polarization with deuterium target, asymmetry with polarized deuterium target, and asymmetry with polarized $^3\text{He}$ target. The data points are taken from Refs.~\cite{Madey:2003av,Geis:2008aa,Glazier:2004ny,Schlimme:2013eoz,Eden:1994ji,Riordan:2010id,Meyerhoff:1994ev,Herberg:1999ud}. For more data points and other theoretical predictions, see Ref.~\cite{Punjabi:2015bba}. The dotted line is the asymptotic prediction $R_n(\infty)=0.864$  with an estimated uncertainty of $\pm 0.11$ for $r=2.08$ in Eq.~(\ref{F1nmod}).} 
\la{fig:Rn}
\end{figure}

\subsection{Holographic predictions for nucleon radii}

We now compute the magnetic root-mean-square (rms) radii of the nucleons from the definition $\langle r_M^2\rangle = -\frac{6}{G_M(0)}\frac{dG_M(Q^2)}{dQ^2}\vert_{Q^2=0}$ and use $\langle r_E^2\rangle=-6\frac{dG_E(Q^2)}{dQ^2}\vert_{Q^2=0}$ to compute the charge mean-square radii of the nucleons. The LFHQCD predictions of different radii are compared with the experimental values in Table~\ref{table:r0}. In determining the charge and magnetic radii, we include the experimental uncertainty by fitting the experimental data and also the systematic uncertainties coming from the LFHQCD model itself. The statistical uncertainties are related to the uncertainties in the probabilities $\gamma_{p,n}$ in the fits of the experimental data with $\chi^2/d.o.f. \sim 0.9$ for different fits. We calculate the systematic uncertainties coming from the inclusion of higher Fock components and the parameter $r$ (only for the neutron Dirac FF) in the FF expressions and also the uncertainty coming form the model as described in the appendix. In all, the radii computed from the LFHQCD model described here are in better agreement with the experimental measurements of all radii where no cancellations of leading terms occur. In particular, the proton charge radius obtained from LFHQCD tends to favor the value obtained from muonic hydrogen Lamb shift experiments (for the most recent experimental values see Ref.~\cite{Pohl1:2016xoo}.   A recent analysis~\cite{Liu:2015jna} of various baryon properties at low $Q^2$ values has been performed in a LFHQCD model where the authors included quark mass in the LFWFs.

\begin{table}[htp]
\begin{center}
\setlength\belowcaptionskip{10pt}
\caption{\label{table:r0} Comparison between the experimental values of the nucleon charge and magnetic radii and LFHQCD predictions from this work. The radii agree with the experimental values~\cite{Agashe:2014kda}. They also agree with the predictions without contributions of higher Fock states made in~\cite{Brodsky:2014yha}.}
\begin{tabular}{|c|c|c|c|c}
\hline
Nucleon radii & Experimental values~\cite{Agashe:2014kda}   & LFHQCD [This work] \\
\hline
$\sqrt{\langle r^p_E\rangle^2}$ &  0.8775(51)\,fm ($ep$ CODATA) &0.801(54)\, fm    \\
\hline
$\sqrt{\langle r^p_E\rangle^2}$ &  0.84087(39) \,fm ($\mu p$ Lamb shift)& 0.801(54)\, fm     \\
\hline
$\sqrt{\langle r^p_M\rangle^2}$ &  0.777(16) \,fm &0.789(79)\, fm     \\
\hline
$\langle(r^n_E)^2\rangle$ &  -0.1161(22) \,$\text{fm}^2$ &  -0.073(30) \,$\text{fm}^2$     \\
\hline
$\sqrt{\langle r^n_M\rangle^2}$ &  0.862(9) \,fm  &0.796(81)\, fm \\
\hline
\end{tabular}
\end{center}
\end{table}

\hspace{5pt}

\subsection{Flavor decomposition of nucleon form factors in light-front holographic QCD}

Recent measurements of neutron form factors made it possible to carry out, for the first time, a flavor separation of the up and down-quark contributions to the nucleon electromagnetic FFs up to $Q^2 = 3.4 ~\text{GeV}^2$~\cite{Cates:2011pz} with results not well understood by existing models. The initial flavor-separation results were later expanded in~\cite{Qattan:2012zf, Qattan:2015qxa}, and have been subject of extensive theoretical analysis with contrasting results, which often show a tension in accounting for the down-quark contribution. Therefore, anticipating the upcoming JLab measurements, we use our present holographic model to compare with existing data and extend our predictions to higher $Q^2$ values. To this end, we compare in Figs. \ref{fig:F1q} and \ref{fig:F2q} the flavor decomposition of various FFs, which follows from the LFHQCD results discussed here, with the experimental results from Ref.~\cite{Qattan:2012zf}. In Fig.~\ref{fig:F2q} the results are scaled by $\chi^{-1}_q$, the limiting values of $F_2^q$ at $Q^2 = 0$, {\it i.e.,} $\chi_u = \mu_u -2 = 1.67$ and $\chi_d = \mu_d -1 = -2.03$. The LFHQCD prediction of a faster increase of the up-quark contribution to $Q^4F_1^u$ for $Q^2>1\,\text{GeV}^2$ compared to $Q^4F_1^d$ is consistent with the flavor decomposition performed in Ref.~\cite{Cates:2011pz}. The flavor decomposed FFs described here are in good agreement with the flavor decomposition which follows from incorporating the Regge contribution into generalized parton distributions~\cite{GonzalezHernandez:2012jv}. A faster falloff of the down-quark contribution with $Q^2$ has been interpreted as a possible axial-vector diquark contribution in Refs.~\cite{Cloet:2008re,Roberts:2007jh,Wilson:2011aa}. Although a complete flavor decomposition requires  a contribution from the strange quark and antiquark, a recent high precision lattice QCD calculation~\cite{Sufian:2016pex} indicates that the strange quark contribution to the proton EMFFs is quite small and becomes even smaller at $Q^2>1\,\text{GeV}^2$.

\begin{figure}[htp]
\begin{center}
\includegraphics[height=7.0cm,width=10.5cm]{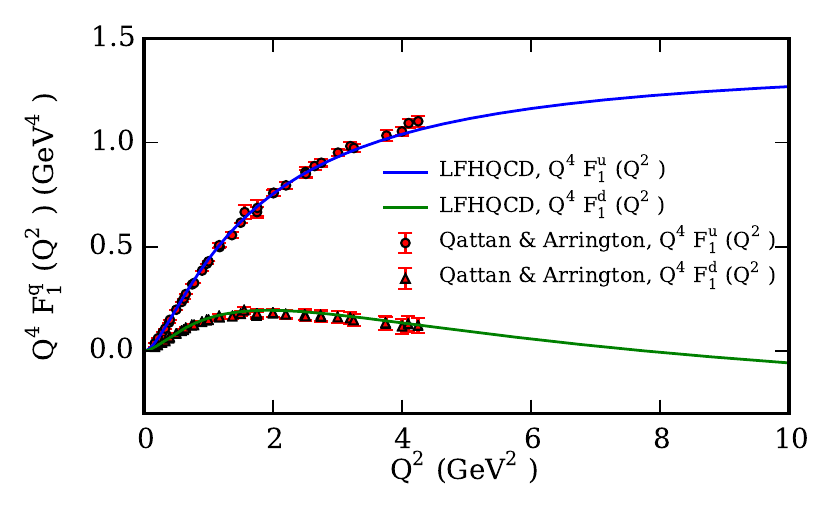}
\end{center}
\setlength\abovecaptionskip{-9pt}
\setlength\belowcaptionskip{-5pt}
\caption{LFHQCD prediction of the up and the down-quark contributions to the Dirac FF multiplied by $Q^4$. The data are from Ref.~\cite{Qattan:2012zf}. }
\la{fig:F1q}
\end{figure}

\begin{figure}
\begin{center}
\includegraphics[height=7.0cm,width=10.5cm]{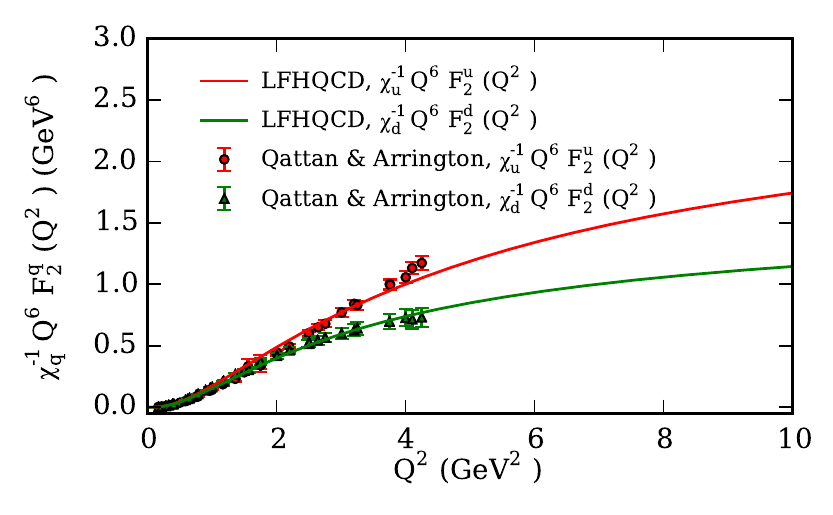}
\end{center}
\setlength\abovecaptionskip{-9pt}
\setlength\belowcaptionskip{-5pt}
\caption{LFHQCD prediction of the up and the down-quark contributions to the Pauli FF multiplied by $\chi^{-1}_q Q^6$. The data are from Ref.~\cite{Qattan:2012zf}.  }
\la{fig:F2q}
\end{figure}

Finally, it is important to recall that we have used a universal value for the confinement scale $\kappa$ in deriving Eq. (\ref{FFtau}), but in fact the value of $\kappa$ for the nucleon wave function, which is obtained from the nucleon slope, is slightly smaller than the value of $\kappa$ in the EM current which is obtained from the rho mass~\cite{Brodsky:2016yod}; it determines the slope of the vector meson trajectory of radial excitations -- the poles in the EM current. Indeed, as explained in the Appendix \ref{appendix}, we have used the difference in the value of the scale $\kappa$, obtained from the average of all meson and all baryon trajectories to evaluate the theoretical uncertainty of our holographic model. Since the wave function determines the low energy bound-state dynamics, we expect that observables which depend on the nucleon wave function, such as radii, are more sensitive to the lower value of $\kappa$, whereas at higher energies, where the amplitudes depend on the structure of the vector meson poles, we would expect that the data are better described by the slightly higher value of $\kappa$ from the rho trajectory of radial excitations. A simple analysis of the data shows that this is indeed the case.

\section{Conclusions} \la{conclusions}

 We have performed a complete analysis of the nucleon electromagnetic form factors in the spacelike region in the framework of light-front holographic QCD. The essential dynamical element in our approach is the embedding of superconformal quantum mechanics in AdS space, which fixes its deformation~\cite{deTeramond:2014asa,Dosch:2015nwa}. The covariant spin structure for the Dirac and Pauli electromagnetic nucleon form factors in the AdS$_5$ semiclassical gravity model encodes the correct power-law scaling for a given twist, ranging from the constituents hard scattering with the photon at high momentum transfer to vector dominance at low $Q^2$. The model also predicts the asymptotic normalization at $Q^2 \to \infty$, which depends on a product of vector meson masses and the valence probability (for the spin-flip Pauli form factor it also depends on the anomalous magnetic moment).
 
 The essential parameter in the model is the confinement scale $\kappa = \sqrt \lambda$ which is universal for the light hadrons and is determined by hadron spectroscopy. This universality holds to better than $10\%$ accuracy~\cite{Brodsky:2016yod}, and has been used to describe a variety of fairly disconnected measurements, such as mass spectra of mesons and nucleons~\cite{Brodsky:2014yha}, form factors~\cite{Brodsky:2014yha} and the infrared behavior of the strong QCD coupling $\alpha_{g1}$~\cite{Deur:2016tte}.

 In the present article, we have considered the effects of the pion cloud which give information on the relevance of higher Fock states. For the spin-flip Pauli form factors, we find an admixture of a five quark state of about 30$\%$ in the proton and about 40$\%$ in the neutron, and essentially no contribution of the higher Fock components to the spin-nonflip Dirac form factors. This relatively important contribution of the higher Fock components to the Pauli form factor of the nucleons is unexpected, and may be related to the fact that the spin-flip form factor corresponds to a change of light-front orbital angular momentum $L = 0 \to L =1$. Likewise, the spin-conserving transition form factor of the proton to a Roper resonance, which can be interpreted as a radial transition from $n = 0 \to n =1$, also requires higher Fock components to describe the low energy data~\cite{deTeramond:2016pov}.

 Since the holographic model does not include spin-flavor structure, we have used the SU(6) symmetry to determine the effective electromagnetic couplings to the quarks for the spin-nonflip form factors. This choice, however, is not precise enough if cancellations of the leading terms are occurring, as in the case of the neutron Dirac form factor. In this case an additional parameter $r$ has to be introduced [see Eq. (\ref{F1nmod})] which accounts for possible SU(6) spin-flavor symmetry breaking effects. For the spin-flip form factors we use the experimental values of the anomalous magnetic moments as an effective coupling. Note that in order to obtain agreement with data, one has to apply a constant shift of the poles predicted by AdS/QCD in the expression for the dressed current to their physical locations. These shifted locations are then obtained from the bound-state equations of the hadrons in this model.

The simple holographic model described here reproduces quite well the main features of the nucleon form factor data. Indeed, with the confinement scale fixed by hadron spectroscopy and the anomalous magnetic moments of proton and neutron fixed by experiment, we have introduced only three free (adjustable) parameters to describe an extensive set of data of the nucleon electromagnetic form factors. Our results for the nucleon form factors and their flavor decomposition agree very well with existing data and provide predictions for the various nucleon form factors in the large momentum transfer regions, which have not been explored by the experiments yet. The charge and magnetic radii of the proton and neutron were extracted and found to agree, within the estimated uncertainty, with their experimental determinations. Our value of the proton charge radius tends to favor the muon Lamb shift determination. In general, the approximations from LFHQCD lead to uncertainties of about $10\%$. Our results should be considered within this typical accuracy. The new JLab experiments will provide a valuable test for our light-front holographic framework which explores the nucleon structure with a minimal number of free parameters.

Since the analytic expression for the form factors (\ref{FFtau}) contains a product of timelike poles, it is especially suited  for also describing the nucleon form factors in the timelike region, as has been done already for the pion form factor in Refs.~\cite{Brodsky:2014yha,deTeramond:2010ez}. The formalism can also be applied to the nucleon transition form factors.

\begin{acknowledgments}

R. S. S. thanks Keh-Fei Liu and Bogdan Wojtsekhowski who provided insight and expertise which assisted this research. S.J.B. is supported by the Department of Energy, Contract No. DE--AC02--76SF00515 and No. SLAC-PUB--16806. A. D. is supported by the U.S. Department of Energy  Office of Science and Office of Nuclear Physics under Contract No. DE--AC05--06OR23177.

\end{acknowledgments}

\appendix 

\section{ESTIMATE OF THE MODEL UNCERTAINTIES} \la{appendix}

Light-front holographic QCD, constrained by superconformal quantum mechanics~\cite{Dosch:2015nwa}, yields a semiclassical description to QCD that can be regarded as a first approximation to full QCD. Therefore, for example, logarithmic terms due to quantum loops are absent in the model. Typically the uncertainties in the spectra are less than 10\%. This is reflected by the fact that the fitted values of the universal confinement scale $\kappa= \sqrt{\lambda}$ differ by about this percentage for the different trajectories~\cite{Brodsky:2016yod}. So we obtain from the the rho trajectory the value $\kappa = 0.537$ GeV, for the nucleon trajectory $\kappa=0.499$ GeV, and from a fit to the rho mass alone $\kappa= m_\rho/\sqrt{2}=0.548$ GeV. Since the rho pole is dominant for the nucleon FFs, we have taken this latter value as the default value in all figures. 

The uncertainties have been estimated in the following way: For the low $Q^2$ region, especially for the charge radii, the form of the nucleon wave function is important; therefore we estimate the uncertainty in this region from the difference of the results obtained with the default value of $\kappa$ and of the result obtained with $\kappa =0.499$ GeV from the nucleon trajectory. For large values of $Q^2$ the FFs are dominated by the product of the rho-meson masses; see Eq. (\ref{asymptotic}). Therefore we estimate the uncertainty for large $Q^2$ by the difference of the default value of $\kappa$ and the value obtained from a fit to all radial and orbital excitations of the rho meson. It typically leads to uncertainties for $F_1$ and $F_2$ below 10\%.

\providecommand{\href}[2]{#2}
\begingroup\raggedright

\endgroup

\end{document}